# LATERAL IBIC CHARACTERIZATION OF SINGLE CRYSTAL SYNTHETIC DIAMOND DETECTORS

A. Lo Giudice[1], P. Olivero[1]*, C. Manfredotti[1], M. Marinelli[2], E. Milani[2],

F. Picollo[1], G. Prestopino[2], A. Re[1], V. Rigato[3], C. Verona[2], G. Verona-Rinati[2], E. Vittone[1]

[1] Experimental Physics Department- NIS Excellence Centre, University of Torino and INFN- sez. Torino, via P.Giuria 1, 10125 Torino, Italy

[2] Dipartimento di Ingegneria Meccanica, Università di Roma "Tor Vergata", Via del Politecnico 1, I-00133, Roma, Italy

[3] INFN- National Laboratories of Legnaro, Viale dell'Università 2, 35020 Legnaro (Pd), Italy

* corresponding author (olivero@to.infn.it)

## ABSTRACT

In order to evaluate the charge collection efficiency (CCE) profile of single-crystal diamond devices based on a p type/intrinsic/metal configuration, a lateral Ion Beam Induced Charge (IBIC) analysis was performed over their cleaved cross sections using a 2 MeV proton microbeam. CCE profiles in the depth direction were extracted from the cross-sectional maps at variable bias voltage. IBIC spectra relevant to the depletion region extending beneath the frontal Schottky electrode show a 100% CCE, with a spectral resolution of about 1.5%. The dependence of the width of the high efficiency region from applied bias voltage allows the constant residual doping concentration of the active region to be evaluated. The region where the electric field is absent shows an exponentially decreasing CCE profile, from which it is possible to estimate the diffusion length of the minority carriers by means of a drift-diffusion model.

## 1. INTRODUCTION

Diamond has extreme electronic and optical properties. The low intrinsic conductivity due to the wide bandgap, the high carrier mobility, the high thermal conductivity, the chemical inertness and the radiation hardness make it a good candidate as particles, UV and x-rays detector in many fields, especially in high radiation environments [1-4]. Although good results were obtained in the past using natural, high pressure high temperature (HPHT) and chemical vapor deposition (CVD) polycrystalline diamond, the last decade witnessed a vast improvement in the diamond detectors performances due to the development of homoepitaxial diamond (single crystal, SC) growth [5]. This material is characterized by high purity and low defect concentration, it exhibits long charge carrier lifetimes, high mobility and does not require a priming procedure before operation [6-9]. SC-diamond detectors exhibited a 100% charge collection efficiency (CCE) and a good energetic resolution. Recent results obtained in the detection of UV light [10], x-rays [11] and neutrons [12] are very promising.

Ion beam induced charge (IBIC) is a very suitable technique to characterise transport properties in wide band gap semiconductors employed as ionizing radiation detectors. Since its development in the early 1990's, IBIC microscopy found widespread applications for the analysis of microelectronic devices, semiconductor radiation detectors, high power transistors, charge-coupled arrays, solar cells, etc [13]. SC-diamond detectors were already studied by means of frontal IBIC [14]. In this paper we report on results obtained in the characterization of diamond SC Shottky diodes by means of lateral IBIC technique. In this configuration the sample is cleaved in order to expose its cross section to the ion beam irradiation. In this way, CCE profiles in the depth direction can be extracted from the cross-sectional maps collected at increasing bias voltage.

## 2. EXPERIMENTAL

The device under analysis was developed starting from a single-crystal diamond grown by CVD technique at the laboratories of Rome "Tor Vergata" University. Diamond was grown on a HPHT substrate in a p-type/intrinsic layered structure by a two-step plasma-enhanced microwave CVD homoepitaxial deposition process described in more details in [8]. A cross-sectional schematic of the device is reported in Fig. 1. A commercial HPHT Ib single crystal diamond $4\times4\times0.4$ mm$^3$ in size was used as substrate. A ~20 μm thick heavily boron-doped (~$10^{20}$ cm$^{-3}$) layer was first deposited on the HPHT substrate followed by the deposition of a 30 μm thick diamond layer diamond layer with a net electrically active acceptor-like defect concentration of the order of $10^{14}$ cm$^{-3}$ [15]. A circular Al contact with a diameter of 2 mm and 200 nm thick was then deposited on the intrinsic diamond surface, while annealed ohmic silver back electrodes were formed on the heavily B-doped layer (see Fig. 1).

The sample was cleaved in order to expose the cross section to the ion beam irradiation and edge-on mounted to perform lateral IBIC experiments. To verify that the good performances of the device were not altered by the cleavage process, a preliminary electrical characterization was performed (see below).

IBIC measurements were carried out at the AN2000 microbeam facility of the National Laboratories of Legnaro (Italy) by using 2 MeV protons. The beam was focused and raster−scanned over the cleaved cross section of the diamond sample from the Schottky contact to the highly doped substrate ($62\times62$ μm$^2$ scan area), as schematically shown in Fig. 1. The beam current was less than $10^3$ protons s$^{-1}$ in order to avoid electronic pile-up and to reduce the radiation damage.

The range of 2 MeV protons in diamond is about 25 μm, as evaluated by the SRIM Monte Carlo Simulation code. Since the electron/hole generation occurs primarily at the Bragg's peak, it is reasonable to assume negligible charge or recombination effects at the irradiated cross-section surface. As a consequence, the induced charge signals collected at the sensitive frontal electrode can be considered as due to the motion of free carriers generated in the bulk of the diamond sample, and subjected to an electric field oriented orthogonally to the two electrodes (i.e. in the y direction, as shown in Fig. 1).

The IBIC signal was acquired using a standard charge sensitive electronic chain (shaping time = 0.5 μs). The calibration procedure was carried following the procedures reported in [16]. The system provided a spectral sensitivity of 270 electrons/channel and a spectral resolution (defined as the FWHM of the 2 MeV proton peak in the Si surface-barrier detector used as reference) of about 3900 electrons, corresponding to an energy resolution of 14 keV in silicon.

## 3. RESULTS AND DISCUSSION

The room temperature current-voltage (I-V) curves in reverse and forward polarizations are reported in Figs. 2a and 2b, respectively. The device exhibits a good rectification effect due to the presence of a Schottky barrier at the Al electrode, whereas the highly doped back contact is assumed to be ohmic.

The linear behaviour of the I-V curve in reverse polarization for V<-20 V indicates the presence of a shunt resistance $R_{sh}=(900\pm6)$ GΩ, as evaluated by the slope of the linear fit (see the inset of Fig. 2a). It is worth remarking that in the reverse bias the cleaved sample showed a leakage current lower than 40 pA at 50 V, thus allowing the acquisition of IBIC signals through a standard charge-sensitive electronic chain with high signal/noise ratio (see below).

The forward bias curve (Fig. 2b) was analysed by the method developed by Cheung and Cheung [17], which is adopted for the analysis of a real diode in presence of a series resistance. The fitting procedure of the IV curve in forward polarization provides an ideality factor n=(1.51±0.04), a series resistance $R_s=(5.1\pm1.6)$ kΩ and a saturation reverse current of $(1.0\pm0.6)\cdot10^{-23}$ A. The resulting fitting curve is shown in Fig. 2b. The high value of $R_s$ is probably due to the highly B-doped electrode at the bottom of the active region. Assuming an effective Richardson constant of 92 A cm$^{-2}$ K$^{-2}$ [18], the barrier height can be estimated around 1.4 eV.

Figure 3 shows lateral IBIC maps at different reverse bias voltages. The charge induced at the sensitive electrode is encoded in colour scale, which represents the median of the IBIC pulse distribution for each pixel. The highest charge is induced by the motion of carriers generated in the proximity of the Schottky electrode, which roughly corresponds to the depletion region where a strong electric field occurs. At zero bias, the maximum induced charge occurs in a region extending around 5 μm beneath the Schottky barrier; as the applied bias increases, the extension of the high efficiency IBIC region widens. Beyond this region, the efficiency monotonically decreases to zero.

A closer inspection of the IBIC spectra at different depths yields additional information on the transport mechanisms responsible of the signal formation. Fig. 4 shows the spectra relevant to the three y positions shown at the top of the maps in Fig. 3 at different bias voltages; for each y position, the IBIC signals reported in the spectra are relevant to the entire scan width (i.e. on the x axis). In correspondence of position 1 (around 2.5 μm beneath the frontal electrode), at zero bias voltage the charge is collected with an average efficiency of 40% . This is the direct evidence of the presence of a built-in depletion region, which allows the the detector to operate in photovoltaic mode [10]. At higher bias voltages, the IBIC spectra show well defined peaks at 100% efficiency with a FWHM of 1.5% corresponding to a spectral resolution of about 30 keV.

At a distance of about 10 μm beneath the Al electrode (position 2), a similar peak is observed for a bias voltage of 60 V; as the bias voltage decreases, the peak broadens and shifts to smaller efficiencies. Finally at about 14 μm depth (position 3), a broad peak at 60% efficiency is observed in correspondence of a bias voltage of 60 V; the electronic threshold, set at about 7% CCE, suppresses most (for $V_{bias}$=25 V) or all (for $V_{bias}$=0) the pulses.

It is worth noticing the similarity between spectra b ($V_{bias}$=25 V) and c ($V_{bias}$=60 V) in Fig. 4. These spectral features are compatible with a pulse formation mechanism not due to carrier drift but to the diffusion of minority carrier (electrons) which are injected into the active region, i.e. the depletion region. Similar spectra have been obtained at different bias voltages (not reported here).

A further confirmation of this fact is clearly highlighted in Fig. 5, where the CCE profiles obtained by projecting the CCE maps along the y axis are reported. These profiles exhibit the same behaviour observed in previously analysed partially depleted silicon p-n junctions [19] and Schottky diodes [20], and interpreted using the drift-diffusion model based on the Shockley-Ramo-Gunn theory [13, 21]. In a simplified version of this model, charge pulses are formed only in regions where an electric field (E) is present, i.e. within the depletion region $W$, and the amount of the induced charge is proportional to the $d/W$ ratio, where $d=\mu \cdot \tau \cdot E$ is the carrier drift length and $\mu \cdot \tau$ is the mobility ($\mu$) – lifetime ($\tau$) product. Due to the fast drift, no recombination occurs during the time of flight of electron and holes throughout the thickness $W$ and the induced charge is equal to the generation charge (i.e. CCE=100%) for carriers generated within the depletion region, which widens as the applied bias voltage increases. Moreover, the dependence of the extension of the depletion region $W$ from the applied voltage $V_a$ can be interpolated with the well known formula:

$$W = \sqrt{2 \cdot \varepsilon_r \cdot \varepsilon_0 \cdot (V_a + V_{bi})/e \cdot N_A} \qquad (1)$$

where $\varepsilon_r$ and $\varepsilon_0$ are the relative and vacuum dielectric constants, $e$ is the elementary electrical charge, $N_A$ is the net electrically active acceptor-like defect concentration and $V_{bi}$ is the built-in potential at the Schottky barrier. As shown in Fig. 6, the experimental data are satisfactorily fitted, yielding a value of the built in voltage of (1.3±0.8) V, to be compared with the value (1.4 V) obtained by C-V measurement, and an estimation of the net electrically active acceptor-like defect concentration of Na = (2.43 ± 0.13)·$10^{14}$ cm$^{-3}$, in good agreement with previous estimations [15].

On the other hand, for *y>W*, i.e. in region where no electric field is present, electrons and holes diffuse away from the generation point. However holes will hardly enter the depletion region because of the opposite electric field, whereas electrons are injected by diffusion into the depletion region and, hence, induce a charge at the sensing electrode. Being diffusion a much slower transport mechanism, recombination effects attenuate the population of electrons entering the active region. As the distance of the generation point from the depletion region increases, assuming an homogeneous material throughout the entire epitaxial layer, the probability of electrons to enter the depletion region exponentially decreases, with a logarithmic slope equal to their diffusion length $L_e = \sqrt{D_e \cdot \tau_e}$, where $D_e$ and $\tau_e$ are the electron diffusivity and lifetime, respectively. The fit procedure of the exponentially decreasing tails reported in Fig. 5 provides an average electron diffusion length of $L_e = (2.6 \pm 0.2)$ μm. This value is in excellent agreement with the value reported in [15] and obtained by x-ray photoconductivity measurements on a SCD Schottky diode based on identically grown intrinsic diamond layer. Using the Einstein relation for the diffusivity/mobility ratio, the product of the zero field mobility ($\mu_e$) and the lifetime is evaluated as $\mu_e \cdot \tau_e = (2.5 \pm 0.3) \cdot 10^{-6}$ V cm$^{-2}$, providing an electron lifetime estimate of ~1 ns, if a mobility of the order of 2000 cm$^2$ V$^{-1}$ s$^{-1}$ is considered [22, 23].

## 4. CONCLUSIONS

IBIC characterization in lateral configuration was proven to be a very suitable technique to study the detection performance and characterize the charge transport properties of SC diamond devices. The detector under investigation presents an electrical rectifying behaviour, due to the Al/diamond Schottky frontal contact. By using 2 MeV proton microbeam, the CCE profile was measured at different applied voltages. In the active region, the device yielded a 100% efficiency with a FWHM of 1.5% corresponding to a spectral resolution of about 35 keV. The width of the region of maximum efficiency increases as the applied bias voltage increases, with a behaviour compatible with that predicted by an ideal Schottky junction with a constant acceptor-like defect concentration in the nominally intrinsic layer $(2.43 \pm 0.13) \cdot 10^{14}$ cm$^{-3}$. The analysis of the exponentially decreasing tails of the CCE profiles occurring in the neutral region highlights an important role played by the diffusion mechanism in the IBIC pulse formation and provides a value of the mobility-lifetime product for electrons of $2.5 \cdot 10^{-6}$ cm$^2$ V$^{-1}$, which corresponds to an average lifetime of the order of ~1 ns if typical mobility values are assumed. These results demonstrate that the electronic properties of the material fully fulfil the requirements for application in ionising radiation detection and dosimetry.

**Acknowledgements**

This work was financially supported in the framework of the INFN experiment DIARAD and MIUR-PRIN2008 National Project "Synthetic single crystal diamond dosimeters for application in clinical radiotherapy". The work of P. Olivero was supported by the "Accademia Nazionale dei Lincei – Compagnia di San Paolo" Nanotechnology grant, which is gratefully acknowledged.

**Figure captions**

Fig. 1. (Color online) Schematics of the diamond detector cross-sectional structure, connection to the acquisition electronic chain, and ion beam probe geometry. The drawing is not to scale.

Fig. 2. (Color online) Reverse (a) and forward (b) current-voltage curves. Experimental data are encoded in black dots, while the red continuous lines represent the fitting curves.

Fig. 3. (Color online) Lateral IBIC maps relevant to bias voltages: a) 0 V, b) 25 V, c) 60 V. The vertical lines at the bottom of the maps indicate the position of the Schottky (S) frontal sensitive electrode and of the B doped back electrode (B).

Fig. 4. (Color online) IBIC spectra relevant to the three y positions indicated by markers in Fig. 3; for each y position, the IBIC signals are relevant to the entire scan width (i.e. on the x axis in Fig. 3).

Fig. 5. (Color online) CCE profiles at different bias voltages obtained by projecting the CCE maps reported in Fig. 3 along the y axis.

Fig. 6. (Color online) Plot of the extension of the depletion region as evaluated from the IBIC maps as a function of applied bias voltage. Experimental data are reported as black dots, while the fitting curve is the red line.

**Figures**

Fig. 1

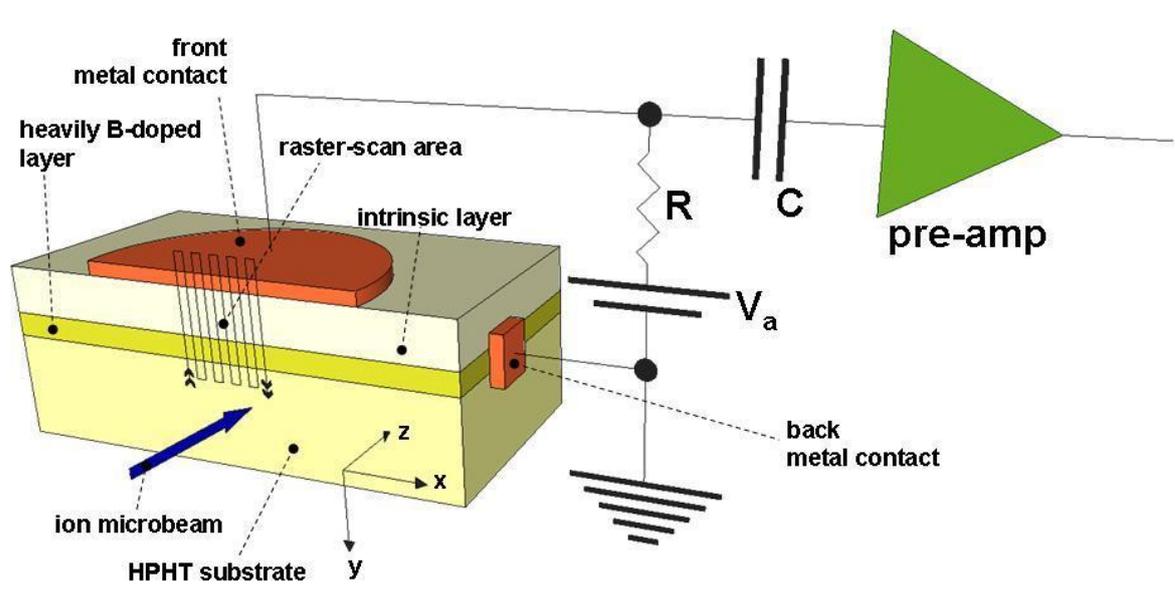

Fig. 2

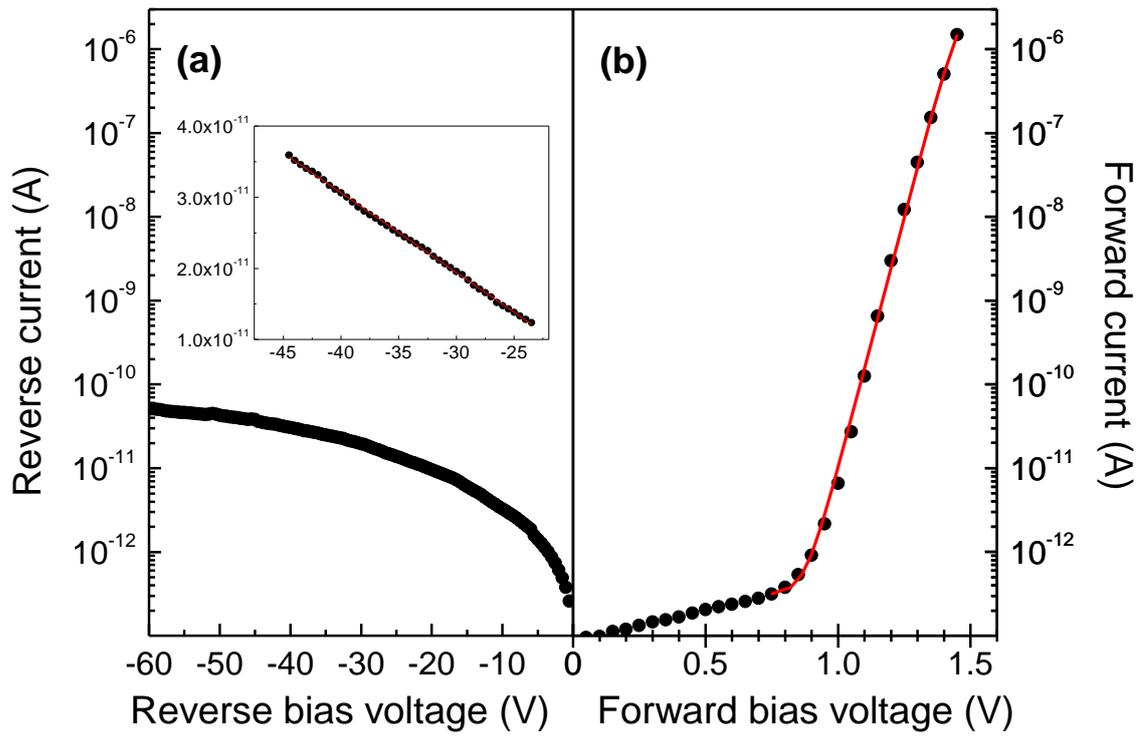

Fig. 3

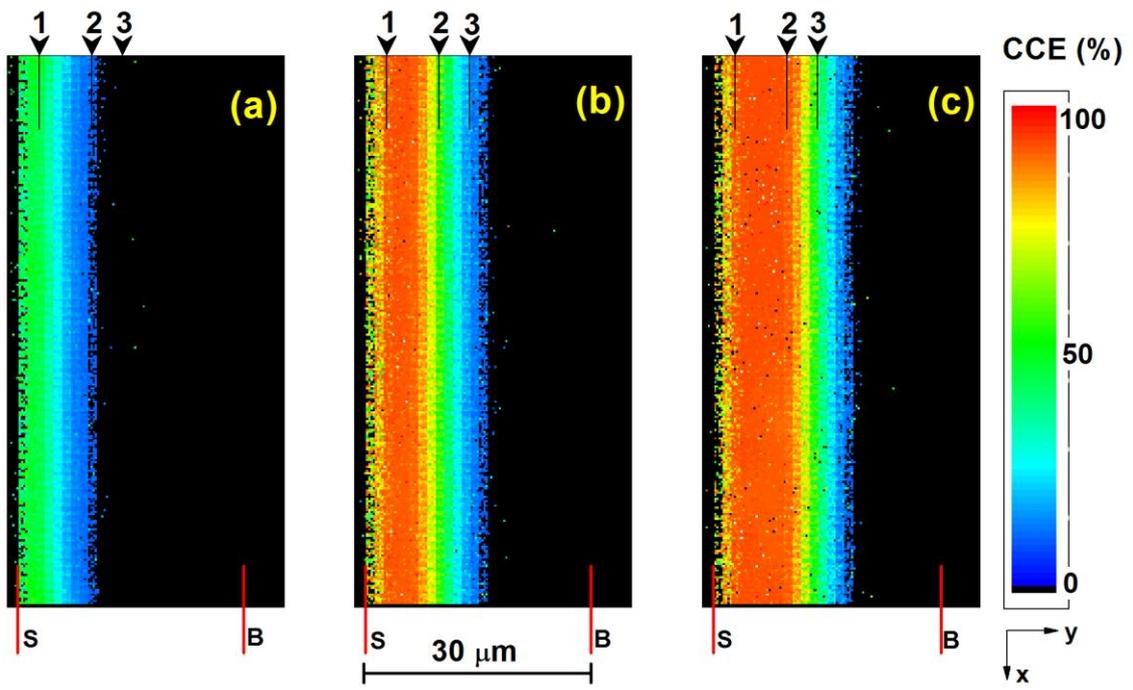

Fig. 4

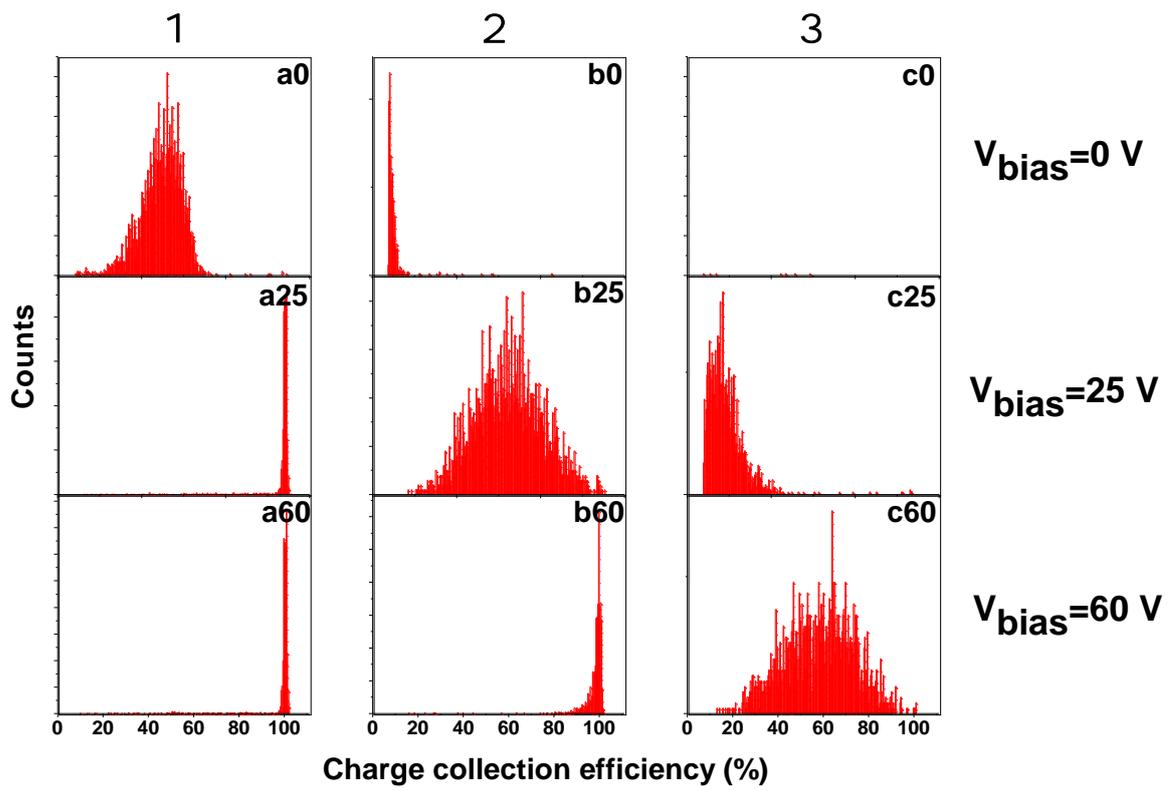

**Charge collection efficiency (%)**

Fig. 5

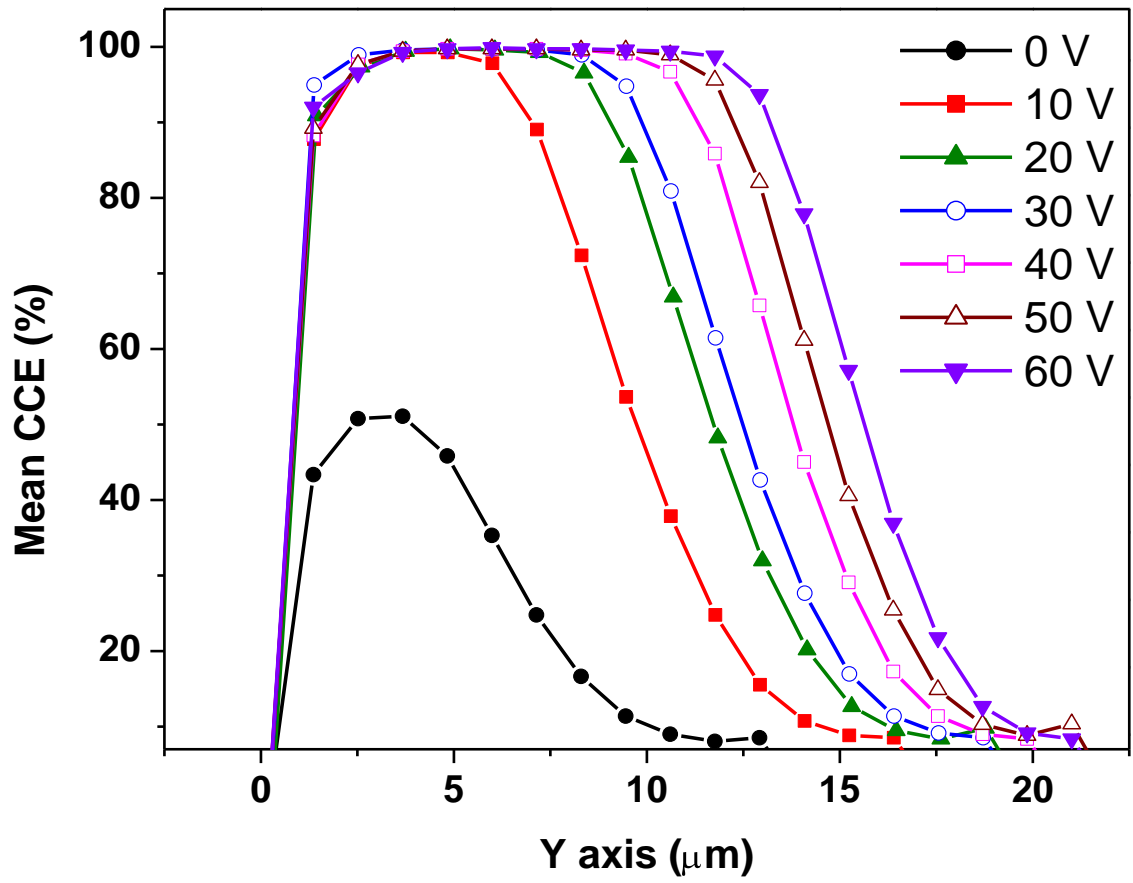

Fig. 6

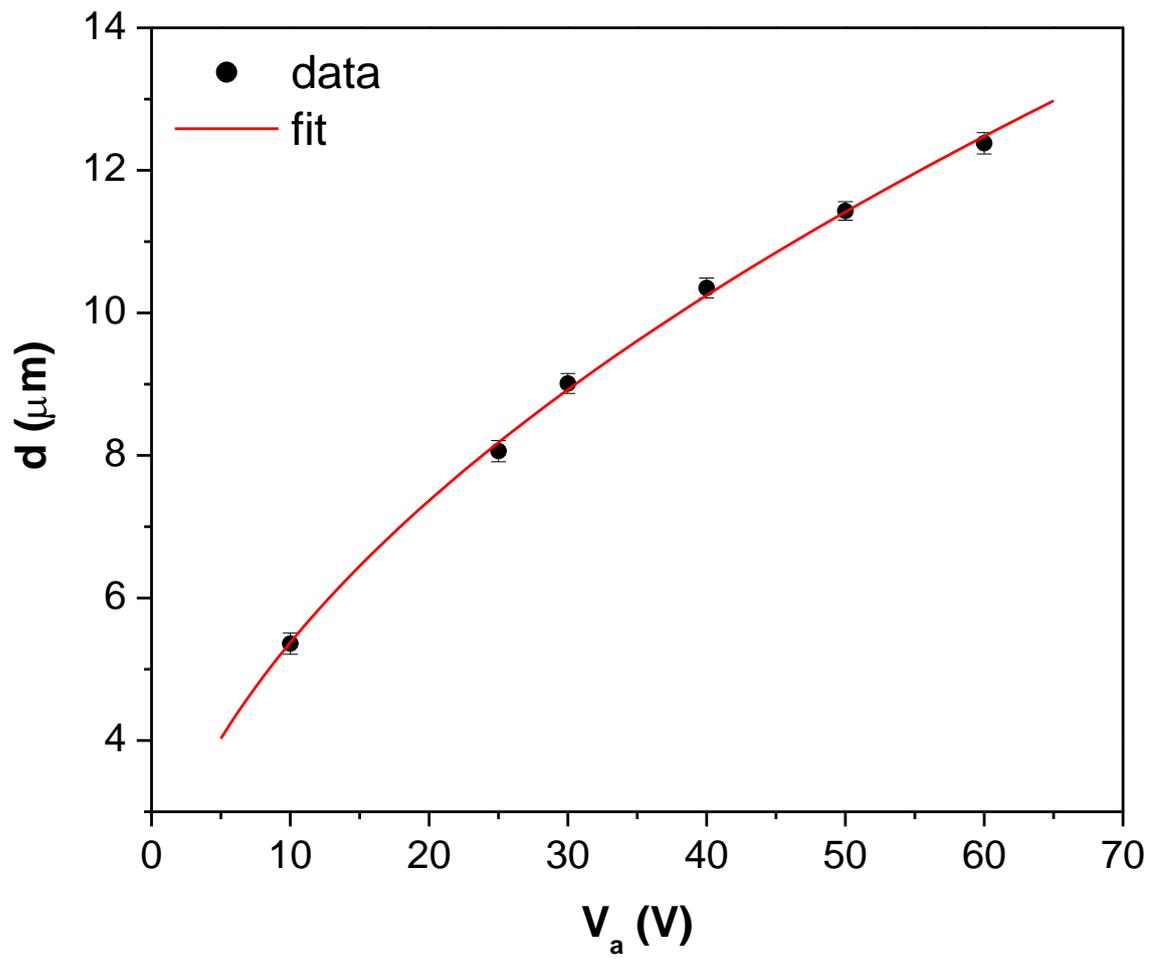